# *Impact of Solar Wind Depression on the Dayside Magnetosphere under Northward Interplanetary Magnetic Field*

S. Baraka [1,2] , L. Ben-Jaffel [3,4]

[1] ***Center for Theoretical and Applied Physics, Al Aqsa University, Gaza, Palestine***
[2] ***Virginia Tech, Blacksburg, VA 24061, resident at National Institute of Aerospace 100 Exploration Way, Hampton, VA 23666***
[3] ***UPMC Paris 06, UMR7095, Institut d'Astrophysique de Paris, F-75014, Paris, France***
[4] ***CNRS, UMR7095, Institut d'Astrophysique de Paris, F-75014, Paris, France***

## Abstract

*We present a follow up study of the sensitivity of the Earth's magnetosphere to solar wind activity using a particles-in-cell model [Baraka and Ben Jaffel, 2007], but here during northward Interplanetary Magnetic Field (IMF). The formation of the magnetospheric cavity and its elongation around the planet is obtained with the classical structure of a magnetosphere with parallel lobes. An impulsive disturbance is then applied to the system by changing the bulk velocity of the solar wind to simulate a decrease in the solar wind dynamic pressure followed by its recovery. In response to the imposed drop in the solar wind velocity, a gap [abrupt depression] in the incoming solar wind plasma appears moving toward the Earth. The gap's size is a ~15 $R_E$ and is comparable to the sizes previously obtained for both $B_z$<0 and $B_z$ =0. During the initial phase of the disturbance along the x-axis, the dayside magnetopause (MP) expands slower than the previous cases of IMF orientations as a result of the abrupt depression. The size of the MP expands nonlinearly due to strengthening of its outer boundary by the northward IMF. Also, during the initial 100 Δt, the MP shrank down from 13.3 $R_E$ to ~9.2 $R_E$ before it started expanding, a phenomenon that was also observed for southern IMF conditions but not during the no IMF case. As soon as they felt the solar wind depression, cusps widened at high altitude while dragged* in an upright position. *For the field's topology*, *the reconnection between magnetospheric and magnetosheath fields is clearly observed in both the northward and southward cusps areas. Also, the tail region in the northward IMF condition is more confined, in contrast to the fishtail-shape obtained in the southward IMF case. An X-point is formed in the tail at ~110 $R_E$ compared to ~103 $R_E$ and ~80 $R_E$ for $B_z$ =0 and $B_z$ <0 respectively. Our findings are consistent with existing reports from many space observatories (Cluster, Geotail, Themis, etc.) for which predictions are proposed to test furthermore our simulation technique.*

**Keywords**: abrupt depression in solar wind flow; dayside magnetopause boundary; northward *IMF*; magnetic reconnection; bow shock; magnetosphere expansion/recovery phase; cusps; particle in cell code.

## 1. Introduction

Chapman and Ferraro [1931 and 1932] were the first to discuss the existence of a boundary to the Earth's magnetic field. In the 1950's the concept of the solar wind was developed. In the early 1960's, Explorer 10 and 12 provided the first measurements of this boundary [Cahill and Amazeen, 1963], called the magnetopause. Since then, we learned that the interaction of the solar wind with the Earth's magnetic field leads to a number of important and intriguing phenomena, many of which are not fully understood because of their complex global nature. These include reconnection at the dayside magnetopause and in the magnetotail, flux transfer events, convection in the magnetosphere/ionosphere and the generation of field-aligned currents [Trattner, K. J. et al, 2008; Imada et al, 2008; Petrinec, S. M. et al, 2003; Nishikawa and Ohtani, 2000; Gosling et al, 1991]. In the resulting magnetospheric configuration, the magnetopause appears as the boundary adjacent to the magnetosphere that is directly exposed to the variable external solar wind.

It follows that by its key position, the magnetopause plays a role of great importance in space physics, since mass, energy, and particles entering the magnetosphere must pass through it [Fairfield et al , 2003 and references within]. Early spacecraft observations of the noon magnetopause position near 10 $R_E$, have shown that the magnetopause boundary is frequently in motion under fluctuating pressure of the Solar Wind [Sonnet et al., 1960; Cahill and Amazeen, 1963; Cahill and Patel, 1967; Kaufmann and Konradi, 1969; Sonnerup, 1987, 1990]. IMF orientation impacts magnetospheric activity, where it minimizes under the northward IMF ($Bz \geq 0$), while it increases with increasing magnitude of southward IMF ($B_z \leq 0$) [Akasofu, 1980].

Most of studies during northward IMF conditions focused on wave generation by solar wind disturbances and resulting propagation in the magnetospheric cavity with the possibility of plasma entry from the magnetosheath through the MP flanks [Terasawa et al., 1997; Sibeck, 2000; Fairfield et al., 2003, 2007, Farrugia et al., 2008, Plaschke et al., 2009]. Usually, the Kelvin-Helmholtz (KH) instability is advocated to explain observation of the MP flanks distortion, a process that may also occur without any solar wind forcing with amplitudes that may exceed a factor of 2, and may lead to plasma entry in the magnetospheric cavity under northward IMF condition [Fairfield et al., 2007]. More recently, the observation and 2D simulation of flux transfer events (FTE) revealed the formation of plasma vortex islands that could be related to time evolving reconnection events near the low-latitude boundary layer (LLBL) [Eriksson et al., 2009]. Also, it is interesting to note recent reports on formation and retreat of many X-lines at high latitudes during a steady northward IMF [Hasegawa et al., 2008, Retinò et al, 2008]. From existing observations and simulations, both the KH instability and reconnection appear as acting processes in the high latitude cusp regions during northward IMF condition, yet their respective efficiency remains uncertain (Frey et al., 2003; Lavraud et al., 2006; Fairfield et al., 2007; Li et al., 2008). More generally, in past studies, the dynamic motion of the MP was mostly considered under the impact of interplanetary shocks that compress the magnetosphere [Zhang et al., 2009]. However, much less attention has been given to study the impact of an abrupt decrease (depression) in the solar wind ram pressure over an extended period, particularly under northward IMF conditions [Baraka and Ben-Jaffel, 2007].

In this paper we present a follow up study to our previous work on the sensitivity of the Earth's magnetosphere to strong variation in solar wind activity, using a particle-in-cell code (PIC) [Baraka, Ben-Jaffel, 2007]. The primarily focus of our study is the response of the dayside MP of the Earth's magnetosphere for which available in situ observations and theoretical simulations are quite abundant [Paschmann et al., 1979; Sonnerrup et al.,1990; Gosling et al.,1991; Baraka and Ben-Jaffel, 2007; Zhang et al., 2009]. This region was also selected because fields and plasmas in the dayside magnetosphere tend to be more ordered, and thereby make field line mapping more tractable. In addition, dayside dynamics tend to be directly driven by solar wind forcing, a property that should assist in studying its sensitivity to the solar wind variations [Murr, 2004].

In our previous work, we used a PIC code to study the interaction of the solar wind with the Earth's magnetosphere during a depression in the solar wind dynamic pressure that was artificially created within the flow of the plasma traveling along the Sun-Earth-direction [Baraka and Ben-Jaffel, 2007]. This effect was applied for two conditions of *IMF*, $B_z$ =0 and $B_z$ <0 respectively. In those studies, during the initial phase of the interaction with the traveling gap, the outer boundary of the magnetopause expanded sunward, which led to the break-up of its structure in the absence of *IMF*, while it sustained its bullet shape in the case of southward *IMF*. Also, in response to the generated gap, low density ensembles of plasma reversed direction and moved sunward against the stream in $B_z$ =0 case. By contrast, this reversal also took place when $B_z$ <0 with much lower density than the case $B_z$ =0. We interpreted this reversal as a result of a gradient in the dynamic pressure from the MP toward the gap region that induced a sunward oriented mechanical force, which stopped these tiny clouds and then reversed their direction of motion [Mishin, 1993; Baraka and Ben-Jaffel, 2007]. We also proposed some similarities between our gap's effect and hot flow anomalies (HFA), particularly plasma flow out of the Sun-Earth line [Sibeck et al 1994, 1999]. Furthermore, the orientation of the cusps was found to be highly affected by the depression in the solar wind flow, while lobes were flared out when $B_z$ =0 due to the applied depression in the solar wind flow. In addition, we found that x lines appeared at different positions in the two *IMF* orientations previously considered. For instance, an x line was found around ~103 $R_E$ for $B_z$ =0, while for $B_z$ <0, its position was found around ~80 $R_E$. These results seem to support the idea that when the *IMF* is non-zero, the x point is more confined toward the planet.

With all these results in hand, it would therefore be useful to see how a depression in the impinging SW could affect the dayside magnetosphere when the *IMF* is included but northward oriented. As in our previous work, we focus on the behavior of the plasma distribution and field's topology at the dayside magnetopause, the formed gap (abrupt depression effect) and the macrostructure of the Earth's magnetosphere. A similar organizational scheme of our previous work will be adopted to easily explore the similarities and the differences in both studies. In addition, a short description of the PIC code by which this study was carried out is presented in section 2 with much more details provided on boundary and initial conditions. For reference, a steady state of the magnetosphere in the presence of northward *IMF* is first obtained with a focus on the kinetic properties of the plasma. Then the SW ram pressure is modified during a short time and the changes in the magnetosphere structure are recorded versus time both for the plasmas and fields distributions. The generation and time evolution of the gap are described before the latter hits the magnetosphere. This description is provided as a reference on the disturbance properties that should help detect similar event in solar wind observations. The time relaxation of the system is then derived through the changes in the magnetopause's sizes following the gap's travel through the entire dayside. Final conclusions are then proposed on similarities and difference with no *IMF* and southward *IMF* cases, respectively.

## 2. Particle-In-Cell Code

### 2.1 Code description: initial conditions

In this study we used 3D, fully electromagnetic and relativistic PIC code, originally written by Buneman et al [1992, 1993 (and the reference therein in Matsumoto and Omura, 1993), 1995]. The original PIC code has been extensively used in the past few years [Nishikawa, 1997, 1998; Nishikawa and Ohtani, 2000; Wodnicka, 2001; Cai et al, 2006], and also with some modifications in terms of handling instabilities and reducing CPU time [Baraka and Ben Jaffel, 2007; Wodnicka, 2009]. In this code, a box, which contains an ensemble of macro-particles, represents the magnetosphere: macro-ions and macro-electrons. Motions of the macro-particles are monitored in 3D under the influence of forces related to electric and magnetic fields through the Lorentz law. The fields themselves are described by Maxwell's equations. The code is relativistic, so no particle's speed can exceed the light's speed c (c=0.5 in the code). We are using the same boundary conditions as initially proposed by Buneman, 1993. Particles leaving the simulation box are assumed to be "absorbed", leaving their charges behind on the boundary. As explained by Buneman (1993), such boundary charging is implemented to reflect the fact that eliminated particles are not too much far from the boundary and that the plasma inside the box keeps "feeling" them. On the average as many ions as electrons escape, the total boundary's charge is moderate. However the immobility of the boundary charges is not real and the surface boundary is made slightly conducting to subtract such surface currents from the surface fields. The same prescription also applies when injecting new particles (for example from the impinging solar wind) from outside. Because the code is charge conserving a difference in particle flux injection is automatically corrected for. To each charged particle injected, a particle of opposite sign is assumed in the same spot. Note that new particles injected from the solar wind on lateral surfaces have a half-space Maxwellian distribution velocity. The boundary conditions for the fields assume no reflection on the box edges as proposed by [Lindman, 1975]. Charge conservation is fulfilled using charge and current deposition scheme as proposed by [Villasenor and Buneman, 1992]. Finally, we remark that the newly injected solar wind's IMF and plasma are assumed to be a frozen in conditions.

Initially, the simulation box is uniformly filled with solar wind plasma composed of pairs of ion-electron macro-particles but no magnetic field is applied. Particles are assumed to have the same speed of Vsw = 0.5*c, where c=0.5 is the light speed in the code. A 3D uniform spatial grid is assumed with a same length scale of $\Delta x = \Delta_y = \Delta_z = \Delta = 1\ R_E$ as in our previous study. A correspondence between the code time unit and the real word time scale is derived here using Courant Condition ($c\Delta t \le \Delta x/\sqrt{3}$) Starting from the solar wind speed assumed in the code Vsw=0.25, we derive that a good time scale could be represented by the Courant time boundary $\Delta t = \frac{\Delta}{c\sqrt{3}} = \frac{1R_E}{2\times V_{sw}\times\sqrt{3}} = \frac{6335}{1732} = 3.6\ \text{sec}$. The significance and conditions of use of this time scaling are further discussed in section 4. Other plasma parameters are scaled in the code such that, $\frac{m_e}{m_i} = \frac{1}{16}$, the solar wind initial electron-proton pair density is $(0.8/\Delta^3)$, the plasma frequencies are $\omega_{e,i} = (0.89, 0.22)$, thermal velocities are $v_{th_{e,i}} = (0.1, 0.025)$, and the Debye length $\lambda_{De,i}$=(0.11, 0.11), where "e"and "i" denotes electron and ion respectively.

It is important to remark that all quantities mentioned above correspond to the initial unperturbed plasma before the application of any magnetic field. These values are also consistent with the fluid description of the solar wind plasma initially injected. However, as we shall see, with the magnetospheric cavity formation, most regions of the magnetosphere will host hotter and/or smaller density plasma. This means the Debye length in those regions will be much larger than assumed for the initial plasma (see section 2.2 for more details).

Besides kinetic effects, it is important to remark that for magnetospheric conditions, the plasma is assumed collisionless. However, when using PIC simulation codes, collision occurs all the time because of the spatial grid used to handle fields and charge and current deposit. In a PIC code, a charge diffuses occupying a volume with a charge distribution defined by a shape function (Langdon, 1971). The numerical collisions lead to the so-called aliasing or grid-heating problem. A lot of work has been done in the past to clarify the onset of the grid heating and its dependency on the plasma parameters (Langdon, 1971; Hockney & Eastwood, 1981 and references therein). From these studies, it appears that the two main parameters are the Debye length ($\lambda_D$) and the plasma frequency ($\omega_p$). In the case of electrostatic simulations (no magnetic field applied) of Maxwellian plasma, Hockney & Eastwood derived a set of $\lambda_D$ and $\omega_p$ values for which grid heating is acceptable. In addition, Langdon (1971) showed that the grid heating occurs by the assumed electrostatic plasma until an asymptotic value of $\lambda_D$ ~0.3 is reached. As shown by Hockney & Eastwood, 1981 (their fig. 9-6 on page 320), our initial and time evolution of ($\lambda_D$, $\omega_p$) ~(0.11, 0.89) are compatible with their optimum path. In addition, we remind that a 3Δx3Δx3Δ shape function of particles is applied on all quantities projected on the grid, a very efficient way to damp instabilities. Obviously, the smoothing is done at the cost of lower spatial resolution, an effect acceptable when dealing with large scales, as it is the case here. We will discuss more in detail this problem in the next section to compare the grid heating derived in this study when a magnetic field is applied.

**2.2 Steady state magnetospheric: numerical stability and kinetic description**

Before the application of any solar wind disturbance, we need to set-up a steady state magnetosphere under northward IMF condition that has the structure expected from past observations. Results are presented in the same way as in our previous study, in order to comparatively conclude on similarities and differences between the current study ($B_z$ >0) and the previous ones ($B_z$ =0 and $B_z$ <0 respectively) [Baraka and Ben-Jaffel, 2007].

The simulation box has the same dimensions of (155,105,105) Δ as in our previous study; where Δ = 1 $R_E$. It is loaded with 2x10$^6$ electron-ion pairs, where Earth is located at (60,52,53) $R_E$. In all graphs, this location is centered at (0, 0, 0) $R_E$, which makes follow up of interpretations more convenient. Starting from a box filled by pairs of electron-ion macro-particles, the Earth's dipole field is switched on, letting the system evolves with time up to 900 $\Delta t$ with an impinging northward IMF component Bz =0.2. The IMF strength is weak but strong enough to provide the expected macrostructure of the magnetosphere after the prescribed elapsed time. Initially, the Debye length is $\lambda_{De,i}$=(0.11, 0.11), respectively for electrons and ions. These values are consistent with the multi-fluid description of the frozen-in magnetized solar wind flow initially assumed. After the planetary and solar magnetic field have been applied, the plasma configuration changed so that it is not clear from the only density distribution shown in Fig. 1 if the PIC code provides a kinetic or fluid description of the particles. To answer this question, we derived a 2D distribution of the Debye length that corresponds to the plasma structure obtained for the steady state configuration displayed in Fig. 1A.

As shown in Fig. 1B, the Debye length contours have a nice distribution that progressively evolve from the multi-fluid description of the impinging solar wind flow into a kinetic description of the magnetospheric plasma. For example, in the magnetosheath nose region, we estimated that the Debye length should be ~0.45. In the current sheet area, the Debye length is found as large as ~1.37. In the cusp regions, we estimate that the Debye length is ~1.17. Finally, in the far tail, we estimate that the Debye length should be around ~0.92. All these values tend to support that our simulation is a kind of hybrid description of the plasma from multi-fluid to kinetic's handling of the particles that is naturally accounted for from the dayside to the nightside of the simulation box. As shown in Fig. 1B, in most of the magnetospheric regions, particles are described kinetically by our PIC code.

Besides kinetic properties, numerical stability of a PIC code should also be addressed. As remarked in section 2.1, grid heating is one of the most discussed issues in PIC simulations. In the case of electrostatic simulations (no magnetic field applied) of Maxwellian plasma, Hockney & Eastwood derived a set of $\lambda_D$ and $\omega_p$ values for which grid heating is acceptable, if not negligible. We may summarize their results remarking that for plasma with $\lambda_D$ < 0.3, grid aliasing should induce a numerical heating that depends on the shape function assumed for the particles on the grid but that should bring the plasma to an asymptotic state with $\lambda_D$ = 0.3-0.5 (Langdon, 1971). According to Hockney, 1971, this artificial heating may corrupt the output of PIC simulation of electrostatic and stable Maxwellian plasma if not handled with care through an adequate selection of the particles shape function. In the present case, we started with a box filled by plasma that has a Debye length that seems rather small (~0.11) but with $\omega_{p,e}\,\Delta t$ ~0.89 and $V_{th,e}$ ~ 0.1. According To Hockeney, 1971, this set of plasma parameters should correspond to stable numerical simulations with a linear shape function for particles as used here. Using the plasma properties in the region upstream of the magnetosphere, we estimated that after 900 time steps, the Debye length is $\lambda_{D,i}$ ~0.14 for ions and $\lambda_{D,e}$ ~0.25 for electrons in the unperturbed solar wind plasma. In addition, numerical heating seems uniformly distributed over the simulation box and smoothly increasing with time until a saturation value $\lambda_{D,e}$ ~0.25 of the Debey length is obtained for electrons. Our results tend to support that the grid aliasing problem is negligible for the heavier ions but follows the same trend for electrons as derived by past studies (Langdon, 1971, Hockney & Eastwood, 1981). For the large structure of the magnetosphere studied here, the numerical noise should not therefore substantially affect our results.

Besides the discussion on the numerical stability of our code, it is also important to compare the magnetosphere so far obtained as a steady state configuration to past observations. In the unperturbed solar wind plasma upstream of the magnetosphere structure, we estimated that the Alfven velocity is $V_A = 0.08c$, the Alfvénic mach number is $M_A \sim 6.1$, and the corresponding magneto-acoustic Mach number is $M_m = 5.2$. The plasma parameter is $\beta \sim 0.35$ with a critical Mach number ~ 2.3 (Edmiston & Kennel, 1984). For this set of plasma parameters, the Earth's bow shock positions along the OX and OY axis are expected from observations at ~(14.8 $R_E$, 29 $R_E$) respectively (Peredo et al., 1995). The magnetopause is expected around ~10.6 $R_E$ along the nose direction (Peredo, 1995; Merka, 2005). It follows that the bow shock to magnetopause standoff distances ratio should be ~1.4 as derived from past observations and models (Fairfield et al., 2001). For comparison, our bow shock positions are ~(15, 29) $R_E$ for sunward and duskward, and the magnetopause nose standoff distance is ~10.8 $R_E$, quite consistently with past models and observations. In addition, the magnetosphere morphology obtained here is also compatible with past observations and models in terms of plasmas and fields distributions, particularly the lobes that are slightly flare out and the fishtail configuration of the magnetotail (Petrinec and Russell, 1996). All these results tend to support that our PIC code when using the appropriate scaled plasma parameters describe quite accurately the larges scales of the magnetosphere, an achievement that provides us with the right tool to investigate in the following the impact of a solar wind disturbance on the steady state configuration obtained at step time 900 $\Delta t$.

### 2.3 Solar wind depression: Gap effect

To obtain a depression in the solar wind flow, Baraka & Ben-Jaffel, 2007, only described the way the disturbance was generated via a sudden change in the incident plasma bulk speed. Although density distributions in the noon-midnight plane clearly showed the gap shape (i.e. Baraka & Ben-Jaffel, 2007, their Fig.4), the description of the resulting structure in the plasma flow upstream of the magnetosphere was not clear. As shown in Figure 2, the exact shape of the disturbance is provided by the normalized plasma parameters across the gap region obtained in the buffer region that is located upstream of the magnetosphere. The shown parameters are measured inside the gap along the Sun-Earth line for each step time. The flow's deceleration resulted in an expansion and cooling of the plasma in the middle of the gap, but produced real shock/discontinuity structures on the edges.

The total pressure curve shows a real solar wind depression. Such input solar wind conditions could be described as a depression consequently followed by compression in the solar wind. It may also be fulfilled by two successive shocks and/or discontinuities that are separated by a lapse of time and that may hit the magnetosphere in a way similar to the gap's effect described here. Such disturbances are common events in the solar wind plasma and field observations from ACE, IMP8, etc.

## 3. Results: Response of the Magnetosphere to a Depression in the Solar Wind Flow

Results are presented in the same way as in our previous study [Baraka and Ben -Jaffel, 2007], in order to comparatively conclude on similarities and differences between the current study ($B_z$ >0) and the previous ones ($B_z$ =0 and $B_z$ <0 respectively) [Baraka and Ben-Jaffel, 2007].

Starting from the steady state configuration obtained at step time 900 $\Delta t$, an abrupt change in the solar wind dynamic pressure was created during 100 $\Delta t$ in order to simulate the variability of the solar wind (Fig. 2). As described in section 2.3, such a reduction in the dynamic pressure resulted in the generation of a gap with two sharp edges ( e.g. [Baraka and Ben Jaffel, 2007]).

### 3.1 Plasma Density Distribution under Northward IMF condition

A snapshot of the system taken at 1001 $\Delta t$ shows the formed gap that appears as a planar structure between x=-47 and x=-32 $R_E$ (see Fig. 3 (A)). The gap approximately has the same size (~ 15 $R_E$) as previously obtained for both $B_z$ <0 and $B_z$ =0. At this time, the magnetopause standoff position is ~ -10.8 $R_E$ (depicted by a vertical dash-dotted line in the figure). A signature of the bow shock can be observed at x ~16 $R_E$ (16 $R_E$ from Earth position). Clouds of plasma in the tail region are filling the formed cavity and are feeding the equatorial plane with a plasma sheet that has a variable thickness along the x-axis. Both cusps are clearly seen, as expected, dayside oriented. More tailward, the lobes are observed almost parallel to the neutral line over 25 $R_E$ from Earth in the nightside; beyond that distance they converge toward the neutral line with denser filling of plasma extending up to 85 $R_E$ (tailward).

In Figure 3(B), taken at 1100 $\Delta t$, the downstream edge of the generated gap is now approaching the dayside magnetopause, where it forms a concave layers over the magnetosheath as a cover. This cover has an apparent thickness along the x-direction of ~5.5 $R_E$. Its concavity can be explained by the incoming plasma, which strongly feels the dipole magnetic pressure on the Sun-Earth line, whilst both ends (toward poles) proceed almost steadily along the x-axis. Inside the gap, clouds of plasma are seen all over and are more uniformly distributed than the previous step time. Apparently, at this particular time the magnetopause standoff position is almost ~-11.5 $R_E$ in the x direction, compared with~13.5 $R_E$ for $B_z$ =0 and ~12.3 $R_E$ for $B_z$ <0. In the night-side, a plasma structure of a relative thickness of ~ 8 $R_E$ is seen starting at x~10 $R_E$. This structure has a ring shape and corresponds to the so-called trapping regions that connect cusps to the equatorial plasma sheet. Both cusps are clearly observed dayside oriented and slightly wider in the poleward direction. At both cusps' peaks, located at (x=-4, z=9) *and (x=-3,z=-7)* $R_E$ respectively for the northern and southern hemispheres, high-density plasma is observed. In addition, other tiny clouds of plasma are observed inside the inner magnetosphere. However, the noise level and the rather moderate number of particles assumed in the simulated plasma makes it difficult to conclude on the origin of these clouds in these particular regions.

The density distribution of the system, taken at step time 1175 $\Delta t$ (75 $\Delta t$ later), shows that the gap is apparently filled by plasma with its downstream edge now almost located along the planet position (0,0,0) (see Fig. 3(C)). By coincidence of the selected time, the new standoff position of the magnetopause at the subsolar point reads ~-18 $R_E$, almost the position of the upstream edge of the gap (see the upright dashed-dotted line in Figure 3(C)). On the contrary, at the same step time, the magnetopause was broken up at ~-15.5 $R_E$ for $B_z$ =0 with a huge cut that opened sunward, while it expanded keeping its classical shape up to ~-17 $R_E$ for $B_z$ <0.

The trapping region is seen as a configuration of clouds of plasma that resembles a ring shape (~7 $R_E$ thick) observed around ~ 15 $R_E$ nightside from Earth (along x-axis). The northward and the southern cusps are dragged in an upright position, but showing a much-extended poleward region compared to a much thinner equatorward region. The thinning of the cusps shape due to the low dynamic pressure of the gap region and its widening by the northward IMF poleward as obtained here confirm and extend previous studies [Burch, 1973; Yamauchi et al., 1996]. Farther away, the cavity structure of the magnetotail can be seen along x out to ~45 $R_E$ (45 $R_E$ from planet). Beyond that distance, in contrast to the cases for $B_z$ =0 and $B_z$ <0, the magnetotail becomes filled up with plasma. Figure 3 (C) also shows a magnetotail shifted toward south and its length shortened, confirming similar results obtained by Nishikawa, Neubert and Buneman [1995].

In panel D of Figure 3, taken at step time 1250 $\Delta t$, the standoff position of the day side magnetopause is now restoring its classical shape at ~-9.5 $R_E$ (comparable to values obtained with $B_z$ =0 and $B_z$<0), yet it appears more confined. The lobes are widely flared out due to the abrupt depression effect that hit their edges. It is remarkably clear that the magnetotail ceased to flare out at 14.4 $R_E$ (~14.4 $R_E$ from the planet position) and restore its standard structure of the lobes/magnetotail during northward *IMF* (see the lobes/magnetotail in figure 1(A) where the gap effect is not yet felt at nightside of the magnetosphere). At this step time, the cusps are dayside oriented but more distorted. A doughnut shape plasma configuration is observed in the nightside cavity, corresponding to the elongated trapping region, previously described, that now lies within the planar gap. Almost 40 $R_E$ nightward, low-density plasma is filling the cavity along z direction.

### 3.2 Magnetic Field Lines Distribution under Northward IMF condition

In the following, we describe the magnetic field's topology corresponding to the plasma density distributions shown in Figure 3. As shown in Fig. 4(A), at step time 1001 $\Delta t$, the northward *IMF* field lines are superimposed to that of the dayside dipole field lines, thusly strengthening them. Accordingly, the magnetic field lines at the magnetopause are closed. In addition, one can see a concave stripe of dense open field lines upstream of the dayside magnetopause (magnetosheath field lines). At the same time, the left side of the simulation box shows distorted field lines (instabilities) that have not been seen when $B_z$ =0, yet they appeared weaker for $B_z$ <0. We are not exactly sure what physical processes are behind these distorted lines, yet it seems related to the presence of the extended planar gap. Indeed, while the steady solar wind flow is clearly super-Alfvénic ($M_A$~6.1), it becomes moderately super-Alfvénic ($M_A$~2.4) in the bottom region (low density region) of the gap. Such conditions of solar wind flow are not unusual as reported by Usmanov et al. [2005]. The contrast between the two adjacent regions could be at the origin of distortions that appear for non-vanishing IMF. It is important to remark that the noise level of the simulation data could also be the origin of the observed distortion of the field lines, since the IMF strength considered is particularly weak in the present study.

At around 44 $R_E$ nightward, the magnetotail is closed like a cocoon. Beyond that distance, field's lines are bent over the cocoon. More interestingly, a double high latitude reconnection (indicated by blue circles) is obtained in the northward and southward cusps areas, confirming both observations and past numerical simulations of the existence of this feature during steady northward IMF [Li et al., 2008; Eriksson et al., 2009]. A zoomed in plot of the encircled regions depicted in Figure 4A, shows that the reconnection between the magnetosheath and magnetosphere fields appears clearly, extending over a few Earth radii and confirming its large-scale structure (see Figure 5). It follows that our simulation PIC code offers an unprecedented opportunity to study in 3D the start-up and time evolution of such reconnection events under northward IMF. Another signature of a large scale reconnection is also visible in the magnetotail region beyond 44 $R_E$ along the x-axis. These results confirm the general configuration that is expected and observed for northward IMF [Burch, 1973; Hasegawa et al., 2008].

In Figure 4(B) taken at 1100 $\Delta t$, as the downstream gap approaches the magnetopause, the magnetosheath field lines are stretched out sunward on the dayside. In addition, the magnetotail cocoon is now closer to the planet. The contraction of the closed structure is another manifestation of reconnection. On the left side of the simulation box, the distorted field lines mentioned in figure 4(A) persist but with a dynamic behavior that could be related both to the instability of the gap region and to statistical noises in the code. As in the previous step time, the double reconnection is still visible at high latitudes in the cusp region, almost at the same positions from Earth.

At step time 1175 $\Delta t$, the position of the downstream boundary of the gap falls together with the inner magnetospheric cavity. Coincidently, the field's lines of the inflated magnetopause are almost tangential to the upstream of the abrupt depression boundary (i.e. Figure 2(C)). The drift in the position of the plasma distribution in the magnetotail to the south, as shown figure 3 (C), is also clearly seen in Fig. 4C for the field lines topology at around 42 $R_E$. The field lines at the cusp region inflated in both northward and southern hemisphere, which matches with the straightening up of the cusps in Figure 3(C). The double high latitude reconnection moved tailward and appears spatially more extended, probably due to the new depressed solar wind conditions that prevail in the gap.

In Figure 4(D), taken at 1250$\Delta t$, as the depression overpasses the planet position, the dayside magnetopause restores its position at around ~-9.5 $R_E$, yet the field line distribution is noisy, if not disturbed, compared to the initial condition before any gap effect was applied. Further to what was observed in Figure 3(D) for the plasma density, the flare out of the fields ceased out after ~ 15 $R_E$ in the nightside. Also, the double high latitude reconnection persists north and south but now appears more disturbed with a probable formation of vortices.

Aside from the large scales changes noticed during the travel of the gap from the day to night side of the magnetosphere, in the following we discuss the expansion/ recovery phases of the magnetopause during the depression/compression of the solar wind dynamic pressure, but this time for northward IMF. With the two cases of vanishing and southward IMF, the following analysis should complete our sensitivity study of the magnetospheric response for the whole range of IMF conditions.

In Figure 6, panels A, B, and C represent the expansion/recovery phase for $B_z$ >0 as measured through the size of the magnetopause along the three main axis. This size is estimated respectively from Earth's position along the x-axis, dawn to dusk for the y-axis, and south to north for the z-axis. We remind that no tilt was assumed for the planet and that x, y, and z represents the Sun-Earth, dawn-dusk, and south-north lines respectively. To locate the magnetopause boundary, we follow the same technique based on the abrupt drop-off of the density by definition of the stagnation region [Baraka and Ben-Jaffel, 2007]. Next, we estimate the position of that edge from the Earth's location.

In panel A of Figure 6, the magnetopause starts expanding from 11 $R_E$ up to 13 $R_E$ (~2 $R_E$ *jump*) between 1099 and 1107 $\Delta t$, then it shrank from 13Re down to 12 $R_E$ between 1107$\Delta t$ and 1115$\Delta t$. Subsequently, the MP linearly expands from the former position up to ~20 $R_E$, a size that it reaches at about 1160 $\Delta t$. Then, apparently, the system relaxes for 6 $\Delta t$, before the MP starts contracting, thusly entering the recovery phase. The later phase starts as soon as the upstream edge of the abrupt depression hits the expanded nose of the magnetopause. The MP then recovers linearly to its original length at 1212 $\Delta t$. It is important to notice that the time variation of the magnetopause's position appears as two distinct phases, one for the expansion and the second for the recovery. This property confirms our claim that the gap disturbance could be sketched as two distinct velocity edges affecting the magnetosphere simultaneously but separately.

In panel B, the magnetopause size in the y-direction shows a different trend. First, the starting position is ~13.3 $R_E$. Second, the magnetopause shrinks during the time interval between 1000 and 1101 $\Delta t$, reaching its minimum size ~9.2 $R_E$ at 1050.78 $\Delta t$. This strange behavior is unique to the dawn-dusk direction but was only observed when IMF (either northward or southern) was added. Further to the process described above, the MP nonlinearly expands from 13.4 to 27.27 $R_E$ between 1101.27 to 1165.7 $\Delta t$. Once the gap effect approaches the magnetopause, the effect of northward *IMF* seemingly hinders the MP expansion/recovery, where the system relaxes during the period of time between 1166 $\Delta t$ and 1223.2 $\Delta t$. Afterward the system recovers very fast as the abrupt depression is over at 1235.7 $\Delta t$.

In panel C, the expansion/recovery of the magnetopause size in shown in the z-direction. In contrast to aforementioned directions, the MP size remains relatively stable at ~ 19 $R_E$ from 1000 $\Delta t$ up to 1136.9 $\Delta t$. Obviously, for that time period, the abrupt depression is still relatively far to be felt by the MP boundary in z-direction. Then, the system expands linearly up to 28.555 $R_E$ at 1190.7 $\Delta t$, before it relaxes over ~37 $\Delta t$ until 1228 $\Delta t$ The recovery phase is linearly attained when the MP reached its initial position (19 $R_E$) at about 1240 $\Delta t$.

## 4. What can we learn about the Magnetosphere under Northward IMF using a PIC code?

For the purpose of consistency with our previous study, in the following, we try to compare our results with existing observations and models, with a focus on the novelty a PIC code brings in understanding the solar wind interaction with the Earth magnetosphere under northward IMF. Here, we stress that the idea of carrying out these case studies is to separate effects (.i.e solar wind parameters, ionosphere, currents systems…etc.) before going into detailed global cases. For example, some principal parameters are not taken into account in our current work, such as ionosphere-magnetosphere coupling, planet tilt, rotational currents, IMF fluctuation values and so on. In our endeavor to carry out these tasks, the dayside magnetosphere (magnetopause) was targeted. In the past, the magnetopause motion was reported often oscillating more or less uniformly in and out a reference position [Formisano et al., 1979] over distances of the order of 0.1 earth radii and with periods of a few minutes [Kaufmann and Konardi, 1969; Sibeck et al., 1991, Safrankova et al., 1997; Haaland et al., 2004; Fairfield et al., 2007; Farrugia et al., 2008; Plaschke et al., 2009, Zhang et al., 2009]. It is clear that when the parameters defining the pressure balance change, the position of the magnetopause will also vary [Semenov et al., 2002].

From this it follows that the primary source of the magnetopause motion is the change in the dynamic pressure of the solar wind that generate waves and instabilities (Kelvin-Helmholtz) that propagate all over the magnetosphere, particularly on the flanks where deformation could be large [Fairfield et al., 2003]. Magnetic reconnection north and south of cusp regions may induce disturbances that could also distort the MP shape. Other processes may trigger magnetopause motion Earthward even when the dynamic pressure is constant, like erosion [Fairfield, 1971, Sibeck, 1991, 1994; Tsyganenko and Sibeck, 1994] or expansion of the MP due to HFA [Sibeck, 1999].

Here, we consider a new process driven by an abrupt depression (gap) in the solar wind ram pressure that we describe and follow in time using a PIC code. As shown in section 2.3, this disturbance could also be realized as two distinct events of the two velocity edges separated by a lapse of time. By contrast with previous cases, the propagation of such feature through the Earth's magnetosphere has never been considered in the past with northward IMF [Baraka and Ben-Jaffel, 2007]. Moreover, the scales of changes in the magnetosphere expected from the depression are far beyond the linear regime of oscillations around a reference structure considered in the past.

### 4.1 Impact of Solar Wind Depression on Large scale structures: the PIC Code Picture

All results so far presented confirm that our updated version of the 3D macro-particle code (PIC) is able to generate the macrostructure of the Earth's magnetosphere and was able to simulate a strong depression of the solar wind under northward IMF conditions, confirming and extending our initial report on vanishing or southward IMF cases. Similarly to our previous study, the applied depression in the SW flow results in the formation of a gap in the plasma flow that has almost the same size of ~15 $R_E$ as obtained for the other IMF conditions. The gap has a planar shape with sharp boundaries perpendicular to the x-axis moving Earthward. In the real solar wind, such depressions with low plasma density on large scales are not unusual events as low-density sub-Alfvénic flows are observed often [Chisham et al., 2000; Usmanov et al., 2005]. In addition, many events in the ACE solar wind data showing two simultaneous velocity peaks separated by a lapse of time may correspond to the disturbance simulated here.

First, we notice a remarkable accumulation of plasma clouds of density above the noise level distributed all along the cavity of the magnetosphere, probably particles diffusing into the magnetosphere, and along the upstream edge of the generated gap as shown in Figure 3. This picture is far from the MHD fluid simulation where the magnetospheric structure appears as smooth boundaries as far as neither diffusion nor reconnection are allowed, unless artificially added. Despite its scaling limitations and the corresponding noise level, our PIC code renders better the kinetic nature of this interaction and may help better understand the formation and evolution of large scale structures of the magnetosphere.

Another large-scale feature of interest in this study is the cusp region with its behavior in response to the solar wind disturbance. Before and after the arrival of the event, the orientation of the cusps was classical and fits rather well with observations (eg Fig. 3) [Burch, 1973; Yamauchi et al., 1995]. However, when the disturbance was right over the planet position covering the whole cusps region, it was remarkable to see how fast the cusps responded to the depression inside the gap and moved anti-sunward, becoming almost in an upright position. Furthermore, they appear very thin in the bottom end at low altitude but much more extended at high altitude, confirming the key role of the solar wind dynamic pressure in controlling the cusps shape and motion [Yamauchi et al., 1995, 1996; Pitout et al., 2006]. In the future, a sensitivity study of the cusps properties in response to the solar wind dynamic pressure for different IMF conditions, particularly when the IMF switches from one direction to another, should provide a reasonable background to understand the main processes that drive their formation and time evolution [Pitout et al., 2009].

One important feature that our PIC simulation obtains naturally is magnetic reconnection, particularly at high latitudes on northward and southern hemispheres. In a recent study, [Lin and Wang, 2006] showed that under a purely northward interplanetary magnetic field (*IMF*), magnetic reconnection in both northward and southern hemispheres leads to a continued formation of newly closed field lines on the dayside, a feature that we observe in our simulation (see Figure 4(A) and 5). Also, recent observations by Cluster and Themis confirm the existence of the double reconnection predicted at high latitudes in north and south cusps areas (indicated by small circles in Figures 4(A)), with persisting X points that our PIC simulations recover nicely as shown in Figure 5 [Hasegawa et al., 2008; Eriksson et al., 2009]. In the same area, the MP flanks deformation due to solar wind disturbances is repeatedly reported [Fairfield et al., 2003, 2007; Farrugia et al., 2008, Zhang et al., 2009, and references within], a deformation that appears naturally in our simulations (eg Figure 3).

All our findings fit rather well with observations, thusly offering a good opportunity to monitor in 3D the high latitude reconnection growth, time evolution, and role with respect to KH instabilities to shape the whole area nearby the MP flanks and cusps [Lund, et al., 2006; Li et al., 2008; Hasegawa et al., 2008; Eriksson et al., 2009].

Farther away in the nightside, using data from ISEE1 and 2, IMP 8 and IMP 7, it was found that the magnetotail ceased to flare during a 12 hours northward *IMF* to an anti-sunward distance of 15 $R_E$ [Chen et al 1993]. As shown in our Figures 4, our results for northward *IMF* seems to agree with these findings, where the magnetotail ceased to flare out after ~14.5 $R_E$. In addition, the magnetotail is seen shifted toward south and its length is shortened. This result is consistent with the occurrence of the magnetic reconnection at the high latitude mantle [Nishikawa et al., 1995, Zheng et al., 2005]. Our PIC code thus offers a great opportunity to study these large-scale features, which in return could help define better the code scaling to the real world.

### 4.2 Time response of the Magnetosphere to Solar Wind Depression: the PIC Code Estimation.

The other useful side of the PIC code is its capability to follow in time the evolving system. However, such a time scale is not clear as far as other key parameters, like the particle masses or number densities, do not correspond to the real magnetosphere. We claimed in section 2 that using the Courant condition may help define a time scale, yet this has to be proven to compare with observations. Here, we stress that our time scale works only for the purpose the code was build for: the formation and evolution of large scales (i.e. macrostructure of the Earth magnetosphere) [Buneman 1993, Nishikawa 1995]. Therefore, as far as the time estimation is used for understanding the evolution of large structures, it should appropriately work. By contrast, it makes no sense to us to use these time scales for the microphysics of very local regions. With that in mind, the analysis of the MP expansion/recovery phases of the MP should provide predictions to test if our assumptions regarding the time scale are acceptable. This comparative exercise may require several studies in order to establish the diagnostic. Nevertheless, we may already notice that the recovery phase always takes place faster that the expansion phase. It is important to remark that the rectangular volume of the gap has a key role in shaping the MP response, as different regions are not hit at the same time. For example, when the gap reaches the MP nose along the x-axis, the other directions do not respond immediately because they do not yet feel the depression. Also, the entire expansion/recovery phases take less than few minutes when using our time scale, which is consistent with most observations on motion and oscillation of the magnetopause that report the same time scale of few minutes [Farrugia et al., 2008, Hasegawa et al., 2008, Plaschke et al., 2009].

In the following, we thus provide more quantitative results that we propose as predictions for the future and when possible to compare to existing observations. It has been noted that the magnetopause expanded non-linearly in all IMF cases along noon, dusk and north direction (x,y, and z). On the other hand, soon after the abrupt depression effect is over, they all recovered almost linearly in 55 $\Delta t$ for the x-axis, 35 $\Delta t$ for the y-axis and 48 $\Delta t$ for the z-axis, respectively. Quantitatively, the magnetopause expansion rate along the x-direction took place with a speed equal to ~ 0.18 in the code units corresponding to an equivalent velocity of 360 km.s$^{-1}$, and recovers at a speed rate equal ~0.16 (~320 km. s$^{-1}$) afterward. Moreover the expansion rate of the MP along the y-direction is 0.177 (~354 km.s$^{-1}$) and the corresponding recovery speed is ~0.31 (~ 620 km.s$^{-1}$). Additionally, the expansion rate as seen in the z-direction showed speed equivalent to 0.12 (~ 240 km. s$^{-1}$), whilst the corresponding recovery speed reaches the value ~0.45 (~900 km.s$^{-1}$). In conclusion, the expansion phase in x-direction is the fastest compared to the expansion in y and z-direction, probably because the subsolar point of the magnetopause is closer to the generated disturbance than the other two boundaries in y and z-directions. On contrary, the recovery phase of the magnetopause in y and z-direction is very much faster than in x-direction. Because the disturbance is effectively generated as a planar cut along x-direction, the elapsed time for the gap to over pass the magnetopause along x-is longer than that elapsed along y and z. This remark is important because it stresses the need to account for a "filling factor" of the disturbance in any analysis of the impact of a strong disturbance (compression or depression) on the whole magnetospheric system. This “filling factor” must be defined through a careful analysis using tools like a PIC code.

For purpose of comparison, we provide hereafter the speed of the expansion/recovery phases obtained in our previous study for $B_z$ =0 and $B_z$ <0. For $B_z$ =0, the MP (expansion/recovery) rates along x are about ~0.13/0.11 (corresponding respectively to 270 and 240 km.s-1), ~ 0.22/0.44 along y (corresponding to 440 and 880 km.s-1 respectively), and ~0.18/0.15 along z (corresponding to 365 and 295 km.s-1 respectively). On the other hand, the MP expansion/recovery along x for $B_z$ <0 is 0.10/0.14 (corresponds to 203 and 283 km.s-1 respectively), and along y 0.164/0.46 (corresponding to 329 and 920 km.s-1 respectively), while along z the values read 0.16/0.33 (corresponds to 316 and 673 km.s-1 respectively). Table 1 summarizes these results and represents a comparative study to the behavior of the magnetopause as it responds to the variability of the solar wind dynamic pressure under different IMF orientations. The 3D and time responses of the MP shown in Table 1 should be compared with observational data in the near future.

## 5. Summary

In this paper, we present a follow up study on the sensitivity of the Earth's magnetosphere to the solar wind activity as initiated by [Baraka and Ben-Jaffel, 2007]. Using a sophisticated PIC code, our goal is to simulate the impact of a strong depression in the solar wind flow on the Earth magnetosphere under northward IMF conditions, thusly completing our previous study for both vanishing and south-oriented *IMF*.

- Our PIC code successfully simulated the macrostructure of the Earth magnetosphere. Based on how the code was built/developed, it offers unprecedented technique to have fields evaluated over the grid nodes, while particles can have any position within the box. Therefore the particles description is hybrid as it gradually evolves from a multi-fluid description in the unperturbed solar wind flow to a kinetic description in the magnetosphere.

- The general properties of a magnetosphere under northward IMF conditions are recovered. Interestingly, the double high latitude reconnection near the cusps is clearly simulated by our PIC code, offering an unprecedented opportunity to study its growth and time evolution in the future.

- One of the findings of this study is that, regardless of the orientation of the IMF, the initial general shape of the solar wind disturbance remains almost unchanged for all IMF values when interacting with the magnetosphere ($B_z$ =0, $B_z$ <0 and $B_z$ >0) . Between the two edges of the disturbance, we noticed that the density of plasma is much larger for $B_z$ >0. Also, more dense plasma populates the cavity at the magnetotail, which results in a southward tail shift (Figure 3(C)); which is in turn a sign of reconnection nearby that region. The nightside magnetosphere is more confined than the other two cases for zero and south *IMF*.

- The recovery phase always takes place separately but faster that the expansion phase. It was also noted that the whole expansion/recovery phases take less than a few minutes when using our time scale, consistently with most observations on the motion and oscillation of the magnetopause that report the same time scale of a few minutes.
- The orientation of the cusp was classical and fits rather well with observations. However, when the depression was right over the planet position covering the whole cusp region, it was remarkable to see how fast the cusps responded to the depression and moved antisunward, becoming almost in upright position. The cusps show a thin end at low altitude and much more extended at high altitude. The effect of the orientation and shape of the cusps for both $B_z$ =0 and $B_z$ <0, is reported stronger than $B_z$ *0.*
- The magnetotail ceased to flare out at $15R_E$ from the planet in the night side, which is in consistent with the observation obtained by ISEE 1 and 2 spacecraft.

**6. Acknowledgements**: The authors warmly thank Prof. K-I Nishikawa for reading the manuscript and for his helpful comments. Work by S. Baraka was supported in part by the Office of the Provost and the Bradley Department of Electrical and Computer Engineering, College of Engineering at Virginia Tech, and, in part, by NASA through grant NNX07AH23G to Virginia Tech. S. Baraka acknowledges the support of CNRS for making the resources of IAP available to the researcher. L. Ben-Jaffel acknowledges support from respectively Centre National d'Etudes Spatiales of France, CNES, under project Inspire, CNRS, and Université Pierre et Marie Curie (UPMC). The authors thank the anonymous reviewer and angeo editorial staff for reviewing our articles.

**Figure 1.A** shows plasma density distribution for unperturbed case taken at 900 $\Delta t$ shown in x-z plane and Earth's position is centered at [0, 0, 0] $R_E$ corresponding to [60, 52, 53] $R_E$ in the simulation box .

**Figure 1.B** shows Debye length distribution in the same plane of density distribution shown in Figure 1A. From magnetosheath to magnetotail, the Debye length in the magnetosphere is ~> 0.5, ensuring a kinetic description of the particles.

**Figure 2** shows the normalize plasma parameters drawn within the gap. This figure is plotted in 100-step time, soon after the disturbance is applied to the system. In five panels the solar parameters are shown in the following order, speed, number density, temperature, magnetic field strength and total pressure. The x axis is reversed time sequence.

**Figure 3**. Time sequence of the response of Earth's magnetosphere to a depression in the incident SW flow for $B_z$ >0. Plasma density is shown in panels A, B, C and D, taken at 1001, 1100, 1175 and 1250 $\Delta t$ respectively. All plots are shown in the x-z plane and centered on Earth's position at [0, 0, 0] $R_E$ corresponding to [60, 52, 53] $R_E$ in the simulation box .

**Figure 4**. Time sequence of the response of Earth's magnetosphere to a depression [abrupt depression effect] in the incident SW flow for $\boldsymbol{B_z}$ >0. Field's lines are shown in panels A, B, C and D, taken at 1001, 1100, 1175 and 1250 $\Delta t$ respectively. All plots are shown in the x-z plane and centered on Earth's position at [0, 0, 0] $R_E$ corresponding to [60, 52, 53] $R_E$ in the simulation box. A zoom of the encircled areas (panel A) will be shown in Figure 3 to highlight the reconnection between the geomagnetic and magnetosheath fields.

**Figure 5:** Zoomed in northward and southward cusp regions showing the reconnection events between geomagnetic and magnetosheath fields as indicated in Figure 2A. **(A)** For the northward cusp, the zoomed area is (0:4, 15:20) $R_E$ from Earth. **(B)** For the southward cusp, the area is (0:4, 13:18) $R_E$ from Earth.

**Figure 6**. Magnetopause's size expansion measured from Earth location [x=60, y=52,z=53] $R_E$ along x, y and z-axis for Bz>0 are shown in panels A, B, and C respectively.

**Table 1**. Summary of the magnetopause variation in 3D due to the depression of the solar wind dynamic pressure. The speed and the time relaxation of the expansion and recovery phases is tabulated for Bz =0 , Bz <0 and Bz >0 respectively

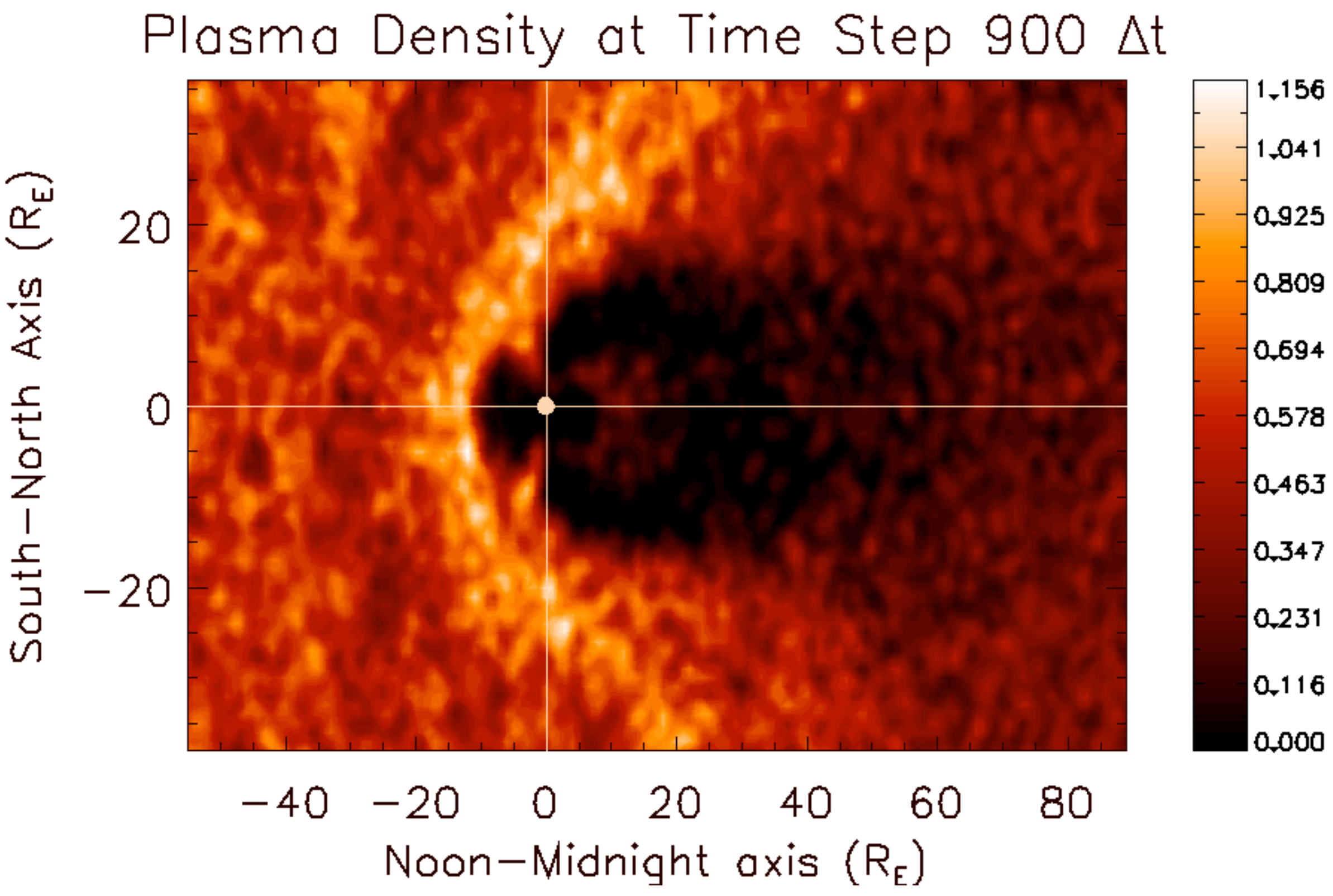

**Figure 1.A**

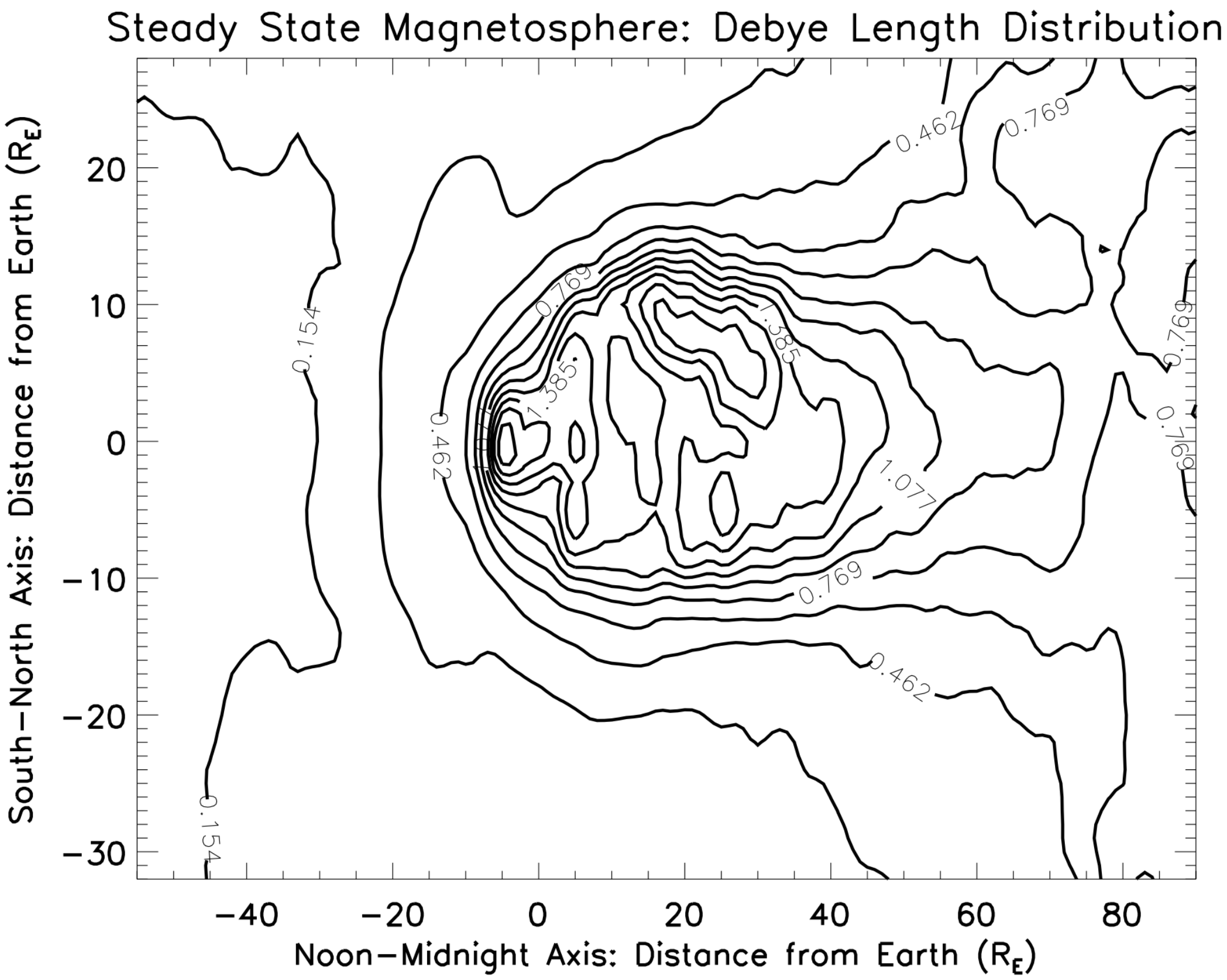

**Figure 1.B**

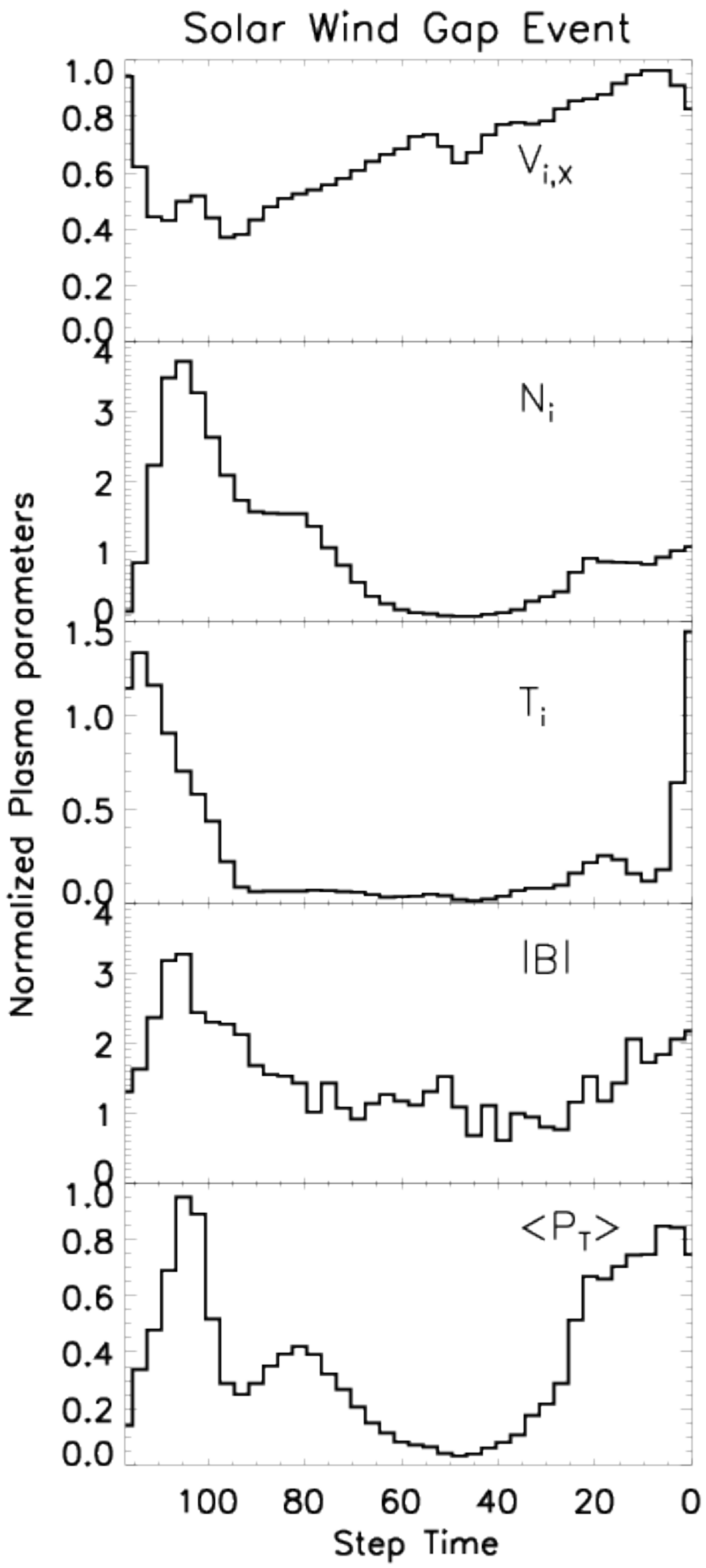
Solar Wind Gap Event
Normalized Plasma parameters
V_i,x
N_i
T_i
|B|
<P_T>
Step Time
1.0
0.8
0.6
0.4
0.2
0.0
4
3
2
1
0
1.5
1.0
0.5
0.0
4
3
2
1
0
1.0
0.8
0.6
0.4
0.2
0.0
100
80
60
40
20
0

**Figure 2**

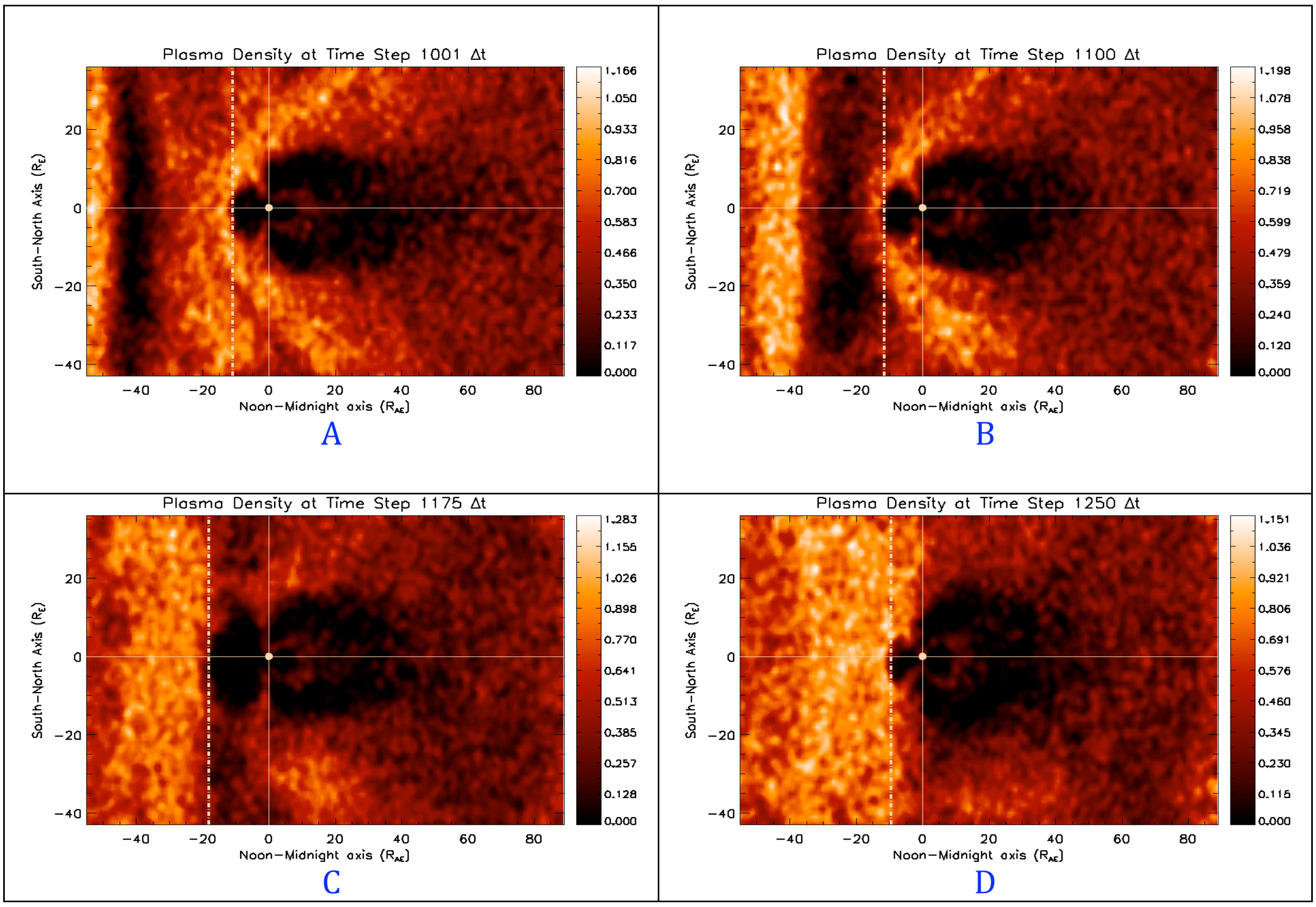

**Figure 3**

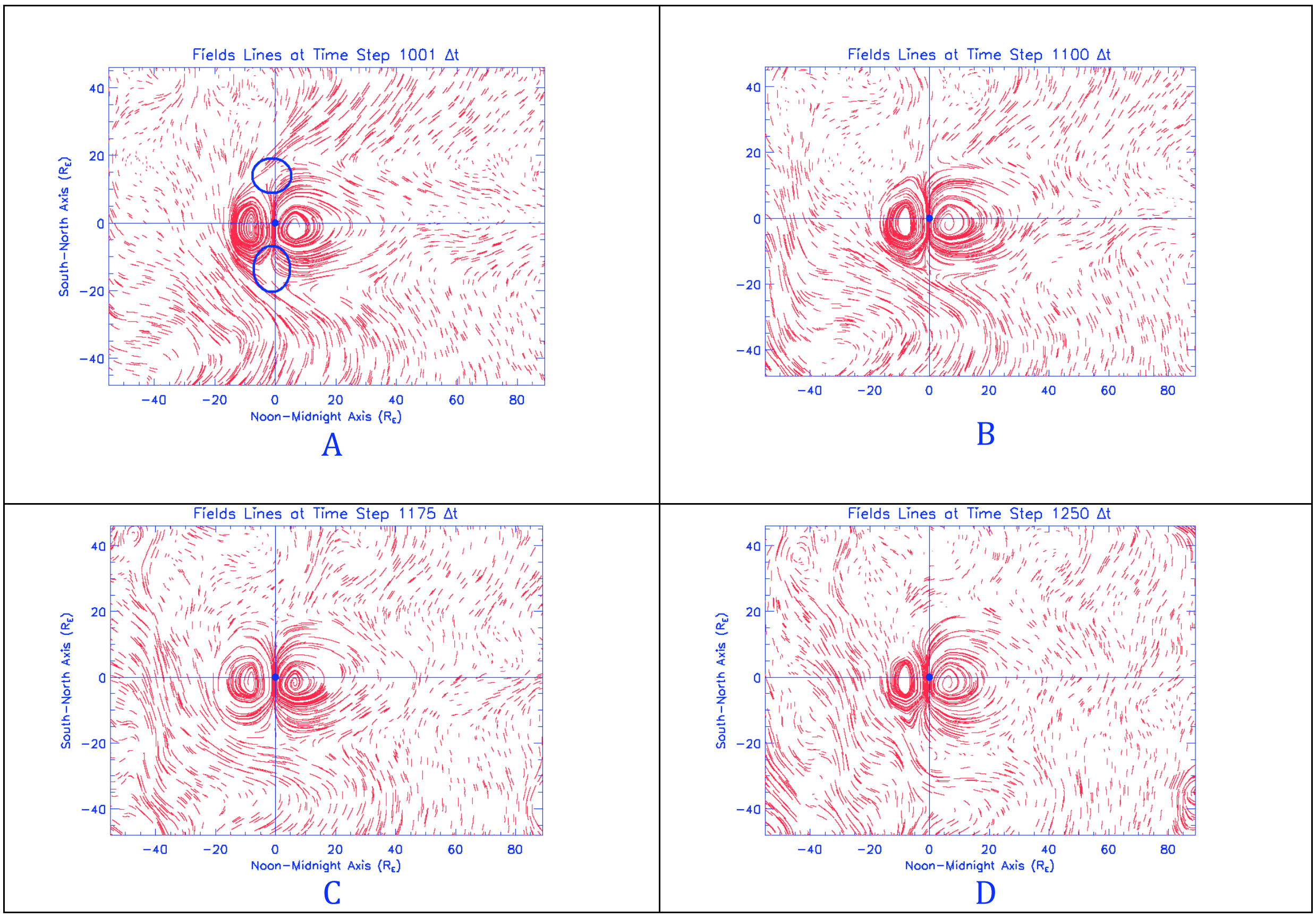
Fields Lines at Time Step 1001 Δt
South−North Axis ($R_E$)
Noon−Midnight Axis ($R_E$)
A
Fields Lines at Time Step 1100 Δt
B
Fields Lines at Time Step 1175 Δt
South−North Axis ($R_E$)
Noon−Midnight Axis ($R_E$)
C
Fields Lines at Time Step 1250 Δt
South−North Axis ($R_E$)
Noon−Midnight Axis ($R_E$)
D

**Figure 4**

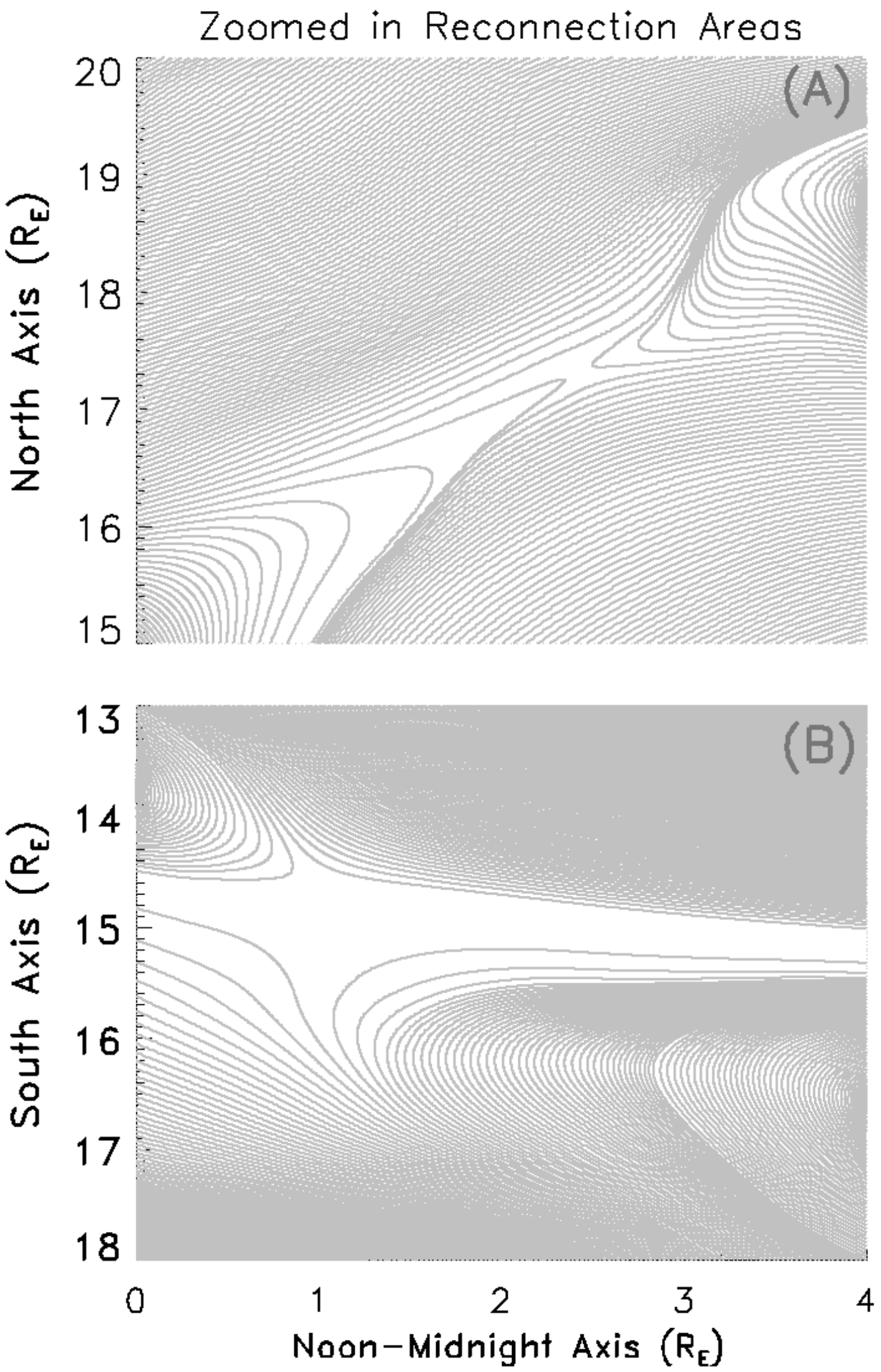

**Figure 5**

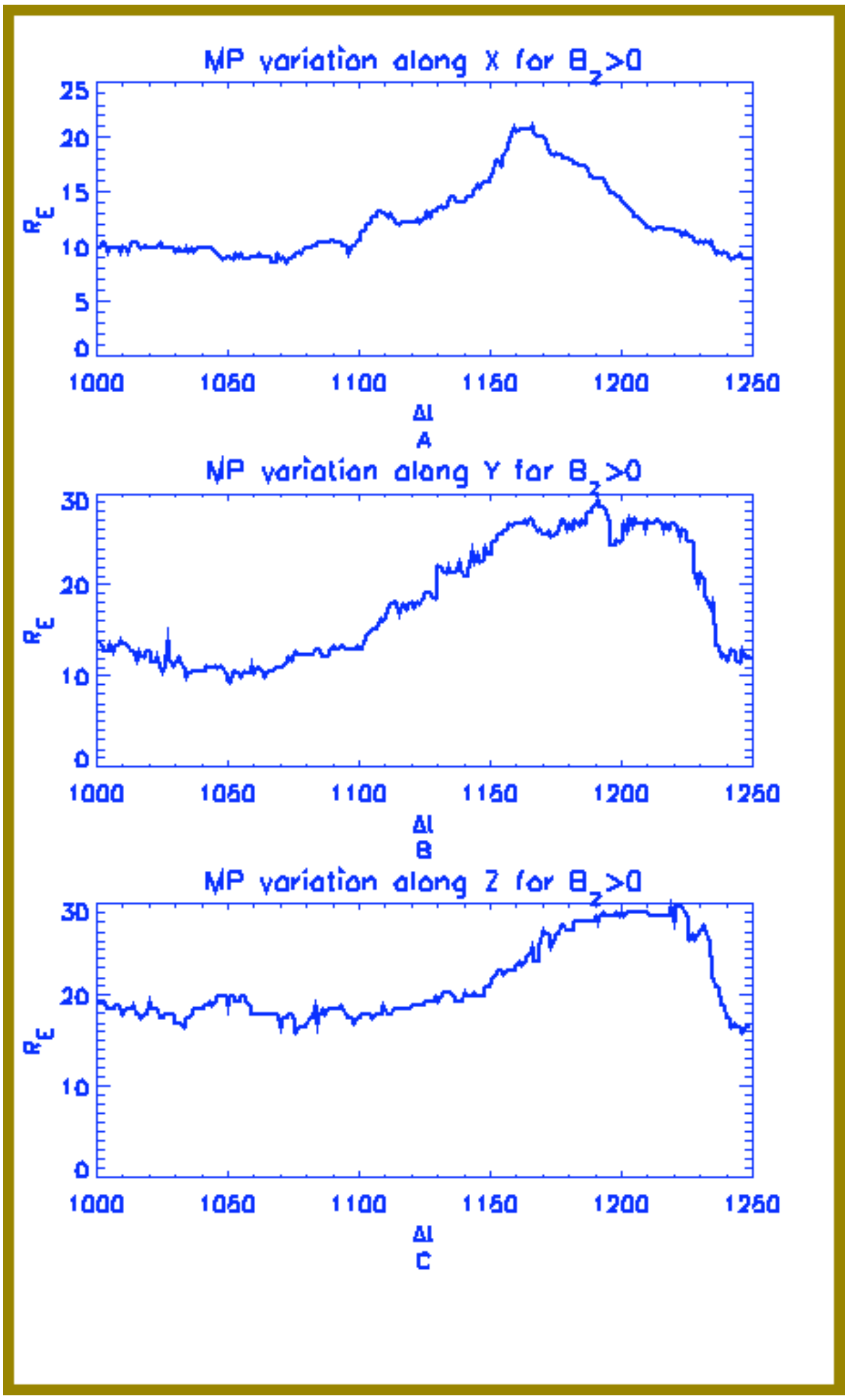
MP variation along X for B_z>0
R_E
ΔI
A
MP variation along Y for B_z>0
R_E
ΔI
B
MP variation along Z for B_z>0
R_E
ΔI
C

**Figure 6**

| ***Case for $B_z$ =0*** | ***Expansion phase of the MP*** | | ***Recovery phase of the MP*** | |
|---|---|---|---|---|
| *Direction* | *Speed [km.s$^{-1}$]* | *Time [minutes]* | *Speed [km.s$^{-1}$]* | *Time [minutes]* |
| X [noon-midnight] | *270* | *3.3* | *237* | *4.2* |
| Y [dawn-dusk] | *441* | *3.9* | *883* | *1.9* |
| Z [south-north] | *365* | *5.1* | *295* | *4.4* |
| ***Case for $B_z$ <0*** | ***Expansion phase of the MP*** | | ***Recovery phase of the MP*** | |
| *Direction* | *Speed [km s$^{-1}$]* | *Time [minutes]* | *Speed [km s$^{-1}$]* | *Time [minutes]* |
| X [noon-midnight] | *203* | *3.5* | *283* | *3.2* |
| Y [dawn-dusk] | *329* | *6.4* | *920* | *2.2* |
| Z [south-north] | *316* | *4.1* | *673* | *2.4* |
| ***Case for $B_z$ >0*** | ***Expansion phase of the MP*** | | ***Recovery phase of the MP*** | |
| *Direction* | *Speed [km.s$^{-1}$]* | *Time [minutes]* | *Speed [km.s$^{-1}$]* | *Time [minutes]* |
| X [noon-midnight] | *360* | *3.8* | *320* | *4.2* |
| Y [dawn-dusk] | *354* | *5.5* | *620* | *1.0* |
| Z [south-north] | *240* | *4.45* | *900* | *1.3* |

**Table 1**